\documentclass{iopconfser}

\usepackage{a4wide,amssymb,cite,graphicx}
\usepackage{epsfig}
\usepackage{hyperref}
\usepackage{amsmath}

\newcommand{\eq}{\mathrm{eq}}
\newcommand{\ie}{\emph{i.e.}}
\newcommand{\eg}{\emph{e.g.}}

\begin{document}

\noindent Conference Proceedings for BCVSPIN 2024: Particle Physics and Cosmology in the Himalayas\\Kathmandu, Nepal, December 9-13, 2024 

\title{Conversion-Driven Freeze-Out: A Common Framework for Dark Matter and Baryogenesis
}

\author{Jan Heisig}

\affil{Institute for Theoretical Particle Physics and Cosmology, RWTH Aachen University,  
D-52056 Aachen, Germany}

\email{heisig@physik.rwth-aachen.de}

\begin{abstract}
We explore dark matter genesis beyond the WIMP paradigm, focusing on the mechanism of conversion-driven freeze-out. This mechanism enables the thermalization of dark matter despite its very weak couplings. While the scenario evades conventional WIMP searches, it predicts novel signatures of long-lived particles at colliders, making it a prime target for upcoming LHC searches. We review various model realizations of this mechanism, highlighting its deep connections to other unresolved problems of the Standard Model. In particular, we show how conversion-driven freeze-out can naturally give rise to baryogenesis, offering a compelling perspective on the origins of both dark and baryonic matter.
\end{abstract}

\section{Introduction}

The nature of dark matter (DM) remains one of the most pressing questions in fundamental physics today.  While the thermal freeze-out of weakly interacting massive particles (WIMPs) holds theoretical appeal, its validity is increasingly questioned due to experimental null results from colliders as well as direct and indirect detection experiments. This has spurred interest in DM production mechanisms that go beyond the WIMP paradigm.

Conversion-driven freeze-out (or coscattering)~\cite{Garny:2017rxs, DAgnolo:2017dbv} has emerged as an intriguing alternative to the standard WIMP freeze-out: while requiring very weak DM couplings that evade the experimental constraints imposed by canonical WIMP searches, it still allows for the thermalization of DM in the early Universe. This makes its predictions insensitive to initial conditions at earlier times, in contrast to other scenarios with very weak couplings -- such as freeze-in or superWIMP production -- that do not provide thermalization.

One of the simplest model realizations of conversion-driven freeze-out is achieved in so-call $t$-channel mediator models where DM state $X$ is accompanied by a slightly heavier mediator particle $Y$ both of which being odd under a new $Z_2$-symmetry stabilizing DM~\cite{LHCDMWGt-ch2025}. Both quark-philic and leptophilic realizations have been explored revealing cosmologically viable scenarios typically pointing toward a DM coupling strength in the ballpark of $10^{-6}$, see \eg~\cite{Garny:2017rxs,Garny:2018icg,Junius:2019dci}. 

In this regime, the rate of annihilation and production of pairs of DM particles in the early Universe is fully negligible, such that DM thermalization and freeze-out proceed via conversion processes only. Those include (inverse) decays $Y\leftrightarrow X i$ and scatterings such as $Yi \leftrightarrow Xj$, where $i,j$ denote SM particles~\cite{Garny:2017rxs}. Effectively, DM dilution proceeds via  $Y\to X$ conversions followed by efficient pair-annihilation of the (color-)charged mediator into SM particles. 

Intriguingly, conversion-driven freeze-out opens up a compelling connection to baryogenesis~\cite{Heisig:2024mwr}: Due to the specific temperature dependence of conversion rates, the departure from equilibrium of DM occurs gradually. It already sets in during the semi-relativistic regime -- before significant DM dilution begins.
This early out-of-equilibrium stage is key to generating a sizeable baryon asymmetry through CP-violating conversions. Notably, the same semi-efficient conversion processes remain large enough to subsequently ensure sufficient DM dilution down to the observed abundance, economically explaining both phenomena through a single interaction type.

While the generation of DM via conversion-driven freeze-out requires a minimal model with two particles only, the simultaneous generation of the baryon asymmetry requires (at least) one additional degree of freedom. One possibility is to introduce $X$ as a multiplet under a global symmetry, as realized in the context of flavored DM~\cite{Agrawal:2011ze}. In fact, the discussed cogenesis mechanism was introduced in Ref.~\cite{Heisig:2024mwr} considering a lepton-flavored DM model, where an initial lepton asymmetry in the SM sector is generated and transferred into a baryon asymmetry via sphaleron processes~\cite{Kuzmin:1985mm}, much like in the well-known thermal leptogenesis scenario~\cite{Fukugita:1986hr}. However, a major difference is that \emph{conversion-driven leptogenesis} does not require lepton-number violation -- sharing similarities to Dirac leptogenesis~\cite{Dick:1999je}. A quark-flavored model yields qualitatively similar results, again requiring no baryon-number violation beyond the SM sphaleron processes, opening up a variety of possible realizations of this idea to be explored. 

While the very weak DM coupling challenges any direct observation of DM-SM interactions, the scenario can be tested through interactions of the mediator with the SM, \eg~through its production at colliders. Notably, due to the small DM coupling, the mediator is long-lived, with a typical decay length in the millimeter-to-meter range, making it a prime target for long-lived particle searches at the LHC~\cite{Heisig:2024xbh}.

In this contribution, we review conversion-driven freeze-out scenarios (Section~\ref{sec:modelreal}), examine their connection to baryogenesis (Section~\ref{sec:baryo}), and outline search strategies at colliders (Section~\ref{sec:collider}). We summarize our findings in Section~\ref{sec:concl}.

\section{Model realizations of conversion-drive freeze-out}\label{sec:modelreal}

Conversion-driven freeze-out can be realized in a variety of new physics models. Arguably, the simplest realization occurs within $t$-channel mediator DM models. These extend the SM by introducing a singlet DM candidate, $X$, and mediator particle, $Y$, both of which being odd under a new $Z_2$-symmetry.  This setup allows for the renormalizable Yukawa-type fermion-portal interaction:
\begin{align}
    \mathcal{L}_{X\hspace{-0.2ex}-Y\hspace{-0.2ex}\mathrm{-SM}} \,=\, & 
  - \lambda \,Y  \,\bar \psi X  \ + \ \text{h.c.} \,.
\end{align}
Here, we assume $X$ to be a Majorana fermion and $Y$ a complex scalar particle while $\psi$ denotes a SM fermion -- a quark or lepton.
Gauge invariance requires the mediator to have the same gauge quantum numbers as the SM fermion $\psi$. Additionally, the scalar mediator can interact via the Higgs portal, leading to the interactions:
\begin{align}
    \label{eq:Y-SM-Lag}\mathcal{L}_{Y\hspace{-0.2ex}\mathrm{-SM}} \,=\, &  (D_{\mu}Y)^\dagger \, D^{\mu}Y 
   -\lambda_H H^{\dagger} H Y^{\dagger} Y\,,
\end{align}
where $D^{\mu}$ is the covariant derivative and $H$ the SM Higgs doublet.

\begin{figure}[t]
  \centering
\includegraphics[width=0.98\textwidth, trim= {0.0cm 0.0cm 0.1cm 0.0cm}, clip]{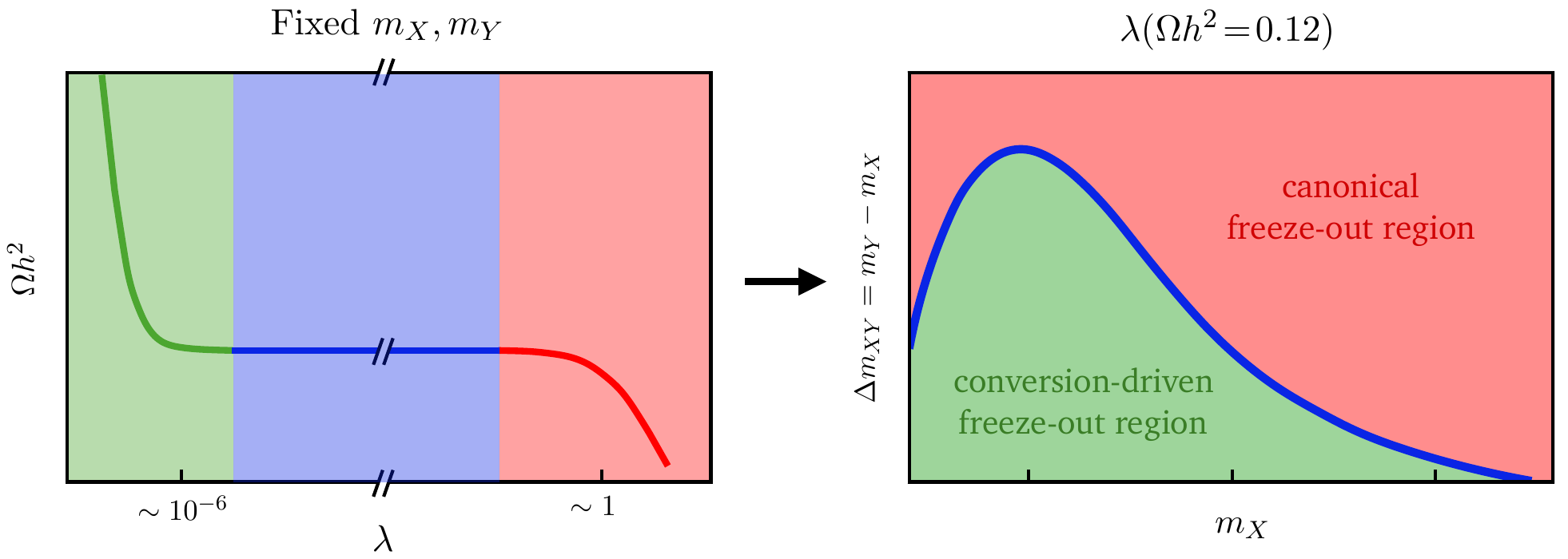} 
\vspace{-1.5ex}
  \caption{
Schematic illustration of the interplay between model parameters and the relic density in different DM freeze-out regimes. Left panel: Dependence of the relic density on $\lambda$ for fixed masses. Right panel: Corresponding freeze-out regions in the $m_X$–$\Delta m_{XY}$ plane, where $\lambda$ is adjusted to satisfy $\Omega h^2 = 0.12$.
}
  \label{fig:schem}
\end{figure}

In the early Universe, several processes govern the evolution of the new physics particles' abundances:
\begin{enumerate}
    \item[\emph{(i)}] Annihilation and coannihilation of DM: These include processes of the form $XX \to \text{SM}$ and $XY \to \text{SM}$, with rates scaling as $\lambda^4$ and $\lambda^2$, respectively.
    \item[\emph{(ii)}] Pair-annihilation of the mediator: The process $YY \to \text{SM}$ contributes to the depletion of $Y$, featuring both a $\lambda$-independent part from the interactions in Eq.~\eqref{eq:Y-SM-Lag} and a $\lambda^4$-dependent contribution (the latter of which is, however, irrelevant in what follows).
    \item[\emph{(iii)}] Conversions between $X$ and $Y$: These occur via (inverse) decays and scatterings, both scaling as $\lambda^2$. 
\end{enumerate}
Depending on the size of $\lambda$, different freeze-out regimes emerge, as illustrated in the schematic plots of Fig.~\ref{fig:schem}. The left panel shows the relic density, $\Omega h^2$, as a function of $\lambda$ for fixed $m_X, m_Y$ with a relatively small mass splitting.
\begin{itemize}
    \item Sizeable $\lambda$: Here, DM annihilation ($XX \to \text{SM}$ and $XY \to \text{SM}$) dominates the depletion of DM, making $\Omega h^2$ inversely proportional to some power of $\lambda$. This corresponds to the canonical WIMP freeze-out regime, highlighted in red.

\item Small $\lambda$: As $\lambda$ decreases, DM annihilation processes become negligible and DM depletion proceeds dominantly via conversion processes ($X \leftrightarrow Y$), followed by efficient mediator pair-annihilation ($YY \to \text{SM}$). The latter is driven by SM gauge interactions (and, for sizable $\lambda_H$, the Higgs portal) and is thus independent of $\lambda$. Although conversion processes scale as $\lambda^2$, they can remain fully efficient despite the small coupling. Unlike DM annihilation, they are not Boltzmann-suppressed ($\propto e^{-m_X/T_\text{fo}}$, with $T_\text{fo}$ the freeze-out temperature) during freeze-out since they do not feature two heavy new physics particles in the initial state. As a result, chemical equilibrium between $X$ and $Y$ is maintained, and the relic density remains virtually independent of $\lambda$ over several orders of magnitude giving rise to the wide plateau in the left plot of Fig.~\ref{fig:schem}. This intermediate regime, occasionally referred to as mediator-annihilation freeze-out, is highlighted in blue.

\item Very small $\lambda$ ($\sim 10^{-6}$): In this regime, DM dilution still proceeds via the combination of conversions and mediator pair-annihilation. However, as $\lambda$ becomes very small, conversion processes become semi-efficient and initiate the chemical decoupling of DM from the mediator (and the SM). This marks the onset of $\lambda$-dependent behavior in $\Omega h^2$, as seen toward the far left of the left plot in Fig.~\ref{fig:schem}. This regime is known as conversion-driven freeze-out~\cite{Garny:2017rxs} (or coscattering~\cite{ DAgnolo:2017dbv}), which is the key mechanism we explore in this article. It is highlighted in green in Fig.~\ref{fig:schem}. 
\end{itemize}

Requiring the predicted relic density to match the observed value, $\Omega h^2 = 0.12$, yields viable solutions in all three regimes within the parameter space of $t$-channel mediator models, as illustrated in the right panel of Fig.\ref{fig:schem}. For large mass splittings $\Delta m$, the plateau in the left panel of Fig.~\ref{fig:schem} lies above $\Omega h^2 = 0.12$ -- indicating that the combination of efficient conversions and mediator annihilation alone is insufficient to dilute DM to the observed level. As a result, sizeable couplings are required, enabling efficient DM annihilation. This corresponds to the canonical freeze-out region, marked in red. Conversely, for small $\Delta m$, the plateau falls below $\Omega h^2 = 0.12$, meaning that a solution exists only in the conversion-driven freeze-out regime, which features very small couplings. This region is highlighted in green in the right panel of Fig.~\ref{fig:schem}. Between these two regions the coupling drops by several orders of magnitude. The corresponding boundary between those regions at which this drop occurs is marked as a blue line in the right panel of Fig.~\ref{fig:schem}. It corresponds to the values of $m_X, m_Y$ for which the plateau in the left panel of Fig.~\ref{fig:schem} coincides with the measured relic density.

The qualitative picture shown in Fig.~\ref{fig:schem} applies for a large class of minimal and non-minimal $t$-channel mediator models. Notably, limited by the size of the mediator-annihilation cross section, the accessible parameter space of conversion-driven freeze-out is constraint toward large $m_X$. For instance, for the bottom-philic minimal $t$-channel model ($\psi=b_\mathrm{R}$)  examined in Ref.~\cite{Garny:2017rxs}, the DM mass is bound to below roughly $4\,$TeV~\cite{Garny:2021qsr}, leaving the scenario exposed to collider probes within the foreseeable future, see Section~\ref{sec:collider} for a discussion of collider constraints.

Interestingly, the small couplings required in conversion-driven freeze-out scenarios ($\lambda\sim 10^{-6}$) provide potential connections to other unresolved issues in the SM\@. For instance, embedding this mechanism within the scotogenic model naturally explains the smallness of active neutrino masses radiatively, while maintaining $\mathcal{O}(1)$ couplings in the scalar potential~\cite{Heeck:2022rep}. Furthermore, the weak thermal contact of DM with the SM bath opens up the possibility of simultaneously generating the baryon asymmetry of the Universe~\cite{Heisig:2024mwr}, a topic we explore in detail in the next section.

\section{Link to Baryogenesis}\label{sec:baryo}

\begin{figure}[t]
  \centering
\includegraphics[width=0.47\textwidth, trim= {0.0cm 0.0cm 0.0cm 0.0cm}, clip]{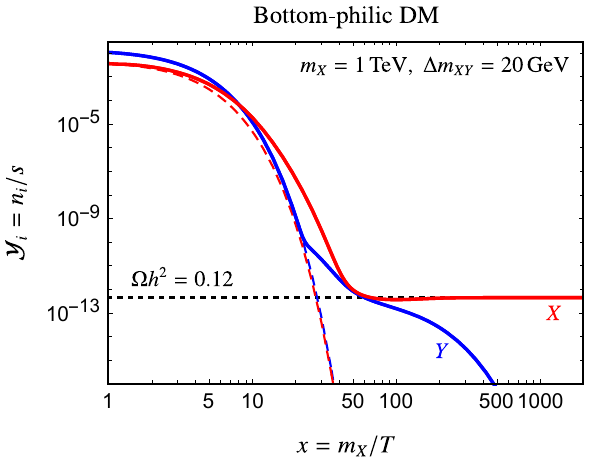} 
\hspace{1.8ex}
\includegraphics[width=0.47\textwidth, trim= {0.0cm 0.0cm 0.0cm 0.0cm}, clip]{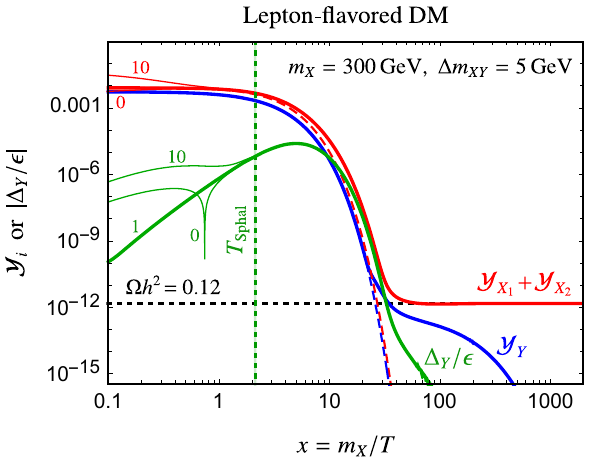} 
\vspace{-1.5ex}
  \caption{
Evolution of abundances, $\mathcal{Y}$, for DM (red) and the mediator (blue) as a function of the temperature parameter $x = m_X/T$ for representative parameter points in the conversion-driven regime, obeying $\Omega h^2 = 0.12$. Left panel: Realization within the bottom-philic minimal model; taken from \cite{Garny:2021qsr}. Right panel: Realization within lepton-flavored DM, which simultaneously explains the baryon asymmetry. The green solid curve shows the mediator asymmetry, $\Delta_Y$, normalized by the $CP$ asymmetry $\epsilon$. The thick solid lines assume the DM multiplet states to be initially equilibrated, while the thin lines labeled ‘0’ and ‘10’ correspond to initial conditions of 0 and 10 times the equilibrium abundances, respectively. The vertical green dashed line marks the sphaleron decoupling temperature; results taken from~\cite{Heisig:2024mwr}.
}
  \label{fig:abund}
\end{figure}

A key aspect of conversion-driven freeze-out is that the conversion rates, $\Gamma_\mathrm{con}$, are semi-efficient. In fact, due to their shallow temperature dependence, $\Gamma_\mathrm{con}$ is kept just above the Hubble expansion rate over an extended period of time, from the relativistic to the non-relativistic regime, thereby supporting thermalization, dilution, and, eventually, chemical decoupling of DM\@. The left panel of Fig.~\ref{fig:abund} shows the resulting evolution of the mediator (blue) and DM (red) abundances, $\mathcal{Y}_Y$ and $\mathcal{Y}_X$, respectively, within the bottom-philic minimal model achieved by solving the respective coupled set of Boltzmann equations~\cite{Garny:2017rxs}. (The plot is taken from Ref.~\cite{Garny:2021qsr}, taking into account important bound-state formation effects of the mediator.) It demonstrates that DM departs from thermal equilibrium already at early times when DM is semi-relativistic. Still, DM is sufficiently strongly coupled to the mediator to successively support significant dilution of the DM abundance in the non-relativistic regime.

Intriguingly, this early departure from thermal equilibrium can be exploited to fulfill the third Sakharov condition for the (simultaneous) generation of a baryon asymmetry~\cite{Heisig:2024mwr}. Since the abundances are still large in the semi-relativistic regime, this asymmetry can be sizeable. However, further necessary conditions for successful baryogenesis must also be met, specifically the presence of $C/CP$ violation and baryon-number violation, \ie~the second and first Sakharov conditions, respectively.

Due to $CPT$ invariance, the total decay rate of $Y$ and $Y^\dagger$ are identical. Hence, the decay $Y\to X \psi$ does not provide any $CP$ asymmetry for the case of a single mediator and DM particle. This leads us to extend the particle content further by at least one additional particle. One possibility is to promote $X$ to a multiplet under a global symmetry with mass eigenstates $X_k$ coupling to the SM fermion flavor $f$ via the coupling matrix $\lambda_{kf}$, as realized in flavored DM~\cite{Agrawal:2011ze}. In this case, the partial width of $Y$ and $Y^\dagger$ for the decay into $X_k$ can differ, enabling a non-vanishing $CP$ asymmetry:
\begin{equation}
    \epsilon_k = \frac{\Gamma_{Y\to X_k \psi}-\Gamma_{Y^\dagger\to X_k \bar\psi} }{\Gamma_{Y\to X_k \psi}+\Gamma_{Y^\dagger\to X_k \bar\psi} }\,,
\end{equation}
where we assume $X$ to be self-conjugate.

The presence of an $CP$ asymmetry can lead to an asymmetry in the mediator particle abundances $\Delta_Y= \mathcal{Y}_Y-\mathcal{Y}_{Y^\dagger}$. The right panel of Fig.~\ref{fig:abund} shows the respective abundances and the mediator asymmetry. Here we consider a leptophilic model, $\psi_f =\ell_{\mathrm{R},f}$, with the minimal working case of two DM flavors coupling to the first two generations only. (The asymmetry is normalized by $\epsilon=\epsilon_1$, which is related to $\epsilon_1$ by $CPT$ invariance.) The asymmetry gradually increases in the semi-relativistic regime due to the departure from equilibrium of the DM multiplet states $X_k$. In fact, the source term for the asymmetry is proportional to $\sum_k\epsilon_k\mathcal{Y}_k/\mathcal{Y}_k^\eq$, requiring not only the departure from equilibrium but also differing $\mathcal{Y}_k$ across the flavors~\cite{Heisig:2024mwr}.

Notably, the result is largely independent of the initial conditions~\cite{Heisig:2024mwr}. As shown by the thin solid lines labeled with `0' and `10', we present the results starting with an initial DM abundance of 0 or 10 times the equilibrium abundance, respectively. Remarkably, in the presence of semi-efficient conversions, the DM abundance still converges to the same out-of-equilibrium abundance regardless of its initial conditions, as already pointed out in Ref.\cite{Garny:2017rxs}. This is directly reflected in the resulting asymmetry, which -- while potentially more sensitive to initial conditions -- is still stable against those conditions in the relevant semi-relativistic regime, \ie~for $m_X/T\gtrsim 2$ in the chosen range. This is a remarkable result and distinguishes conversion-driven freeze-out from other scenarios of cogenesis that explore freeze-in.

Finally, we address the first Sakharov condition. Notably, within the discussed scenario, we do not (need to) introduce baryon or lepton number violation in the new physics sector. Accordingly, the conversion processes induce an asymmetry in the SM lepton sector of equal magnitude but opposite sign to that of the mediator. However, unlike the mediator asymmetry, the lepton asymmetry is transferred to a baryon asymmetry via sphaleron processes, which remain efficient until a temperature of approximately $130$ GeV. At this point, sphalerons decouple, freezing in the baryon asymmetry. Crucially, this decoupling occurs before the rapid decline of the mediator abundance and, consequently, its asymmetry at $m_X/T\gtrsim 5$, making conversion-driven freeze-out with $m_X$ in the range of hundreds of GeV an efficient framework for the cogenesis of DM and the baryon asymmetry.

The considered lepton-flavored model can explain both observations in the resonant regime of quasi-degenerate DM multiplet states and a relatively small mass splitting $\Delta m_{XY}$ for DM masses up to around 500\,GeV~\cite{Heisig:2024mwr}. This parameter region offers testable predictions for collider searches.

\section{Tests at colliders}\label{sec:collider}

In the scenarios discussed above, the mediator is electrically or color-charged, leading to a sizeable production cross-section at colliders. Interestingly, and quite generally for conversion-driven freeze-out, the mediator exhibits macroscopic lifetimes, giving rise to prominent long-lived particle signatures. This can be understood from the following rough estimate. For conversion processes to initiate freeze-out, their rate must be comparable to the Hubble expansion rate, $\Gamma_\mathrm{con} \sim \mathcal{H}$. Since conversion comprises both decays and scatterings, the former requires $\Gamma_\mathrm{decay} \lesssim \mathcal{H}$, implying a corresponding decay length of $c\tau_Y \gtrsim\mathcal{H}^{-1}$. This translates to a characteristic scale in the ballpark of decimeters for typical freeze-out temperatures for $m_X\sim \mathcal{O}(100\,\mathrm{GeV})$. A more precise calculation reveals that $c\tau_Y$ typically ranges from millimeters to meters if $\Gamma_\mathrm{decay} \sim \Gamma_\mathrm{con}$\cite{Garny:2017rxs,Heisig:2024xbh}, while it can be much larger if conversion is dominated by scatterings, $\Gamma_\mathrm{decay} \ll \Gamma_\mathrm{con}$~\cite{Garny:2018icg}.

This lifetime range can only be probed using a combination of different search strategies~\cite{Heisig:2024xbh}. For large decay lengths well above a meter, a significant fraction of mediators traverse the entire detector and can be identified through anomalous ionization losses and time-of-flight measurements in heavy stable charged particle searches. For $10\,\mathrm{cm} \lesssim c\tau_Y \lesssim 1\,\mathrm{m}$, searches for disappearing tracks -- where the charged mediator decays into DM and soft visible objects -- are promising. For even shorter lifetimes, the mediator track might not be reconstructed, necessitating searches for displaced objects such as displaced jets or leptons from the mediator's decay. Finally, missing energy searches, which do not specifically target the long-lived nature of the mediator, may also provide sensitivity to the scenario, though their reach is naturally weaker since they do not exploit the distinct long-lived particle signature.

\begin{figure}[t] 
\centering 
\includegraphics[width=0.46\textwidth, trim= {0.0cm 0.0cm 0.0cm 0.0cm}, clip]{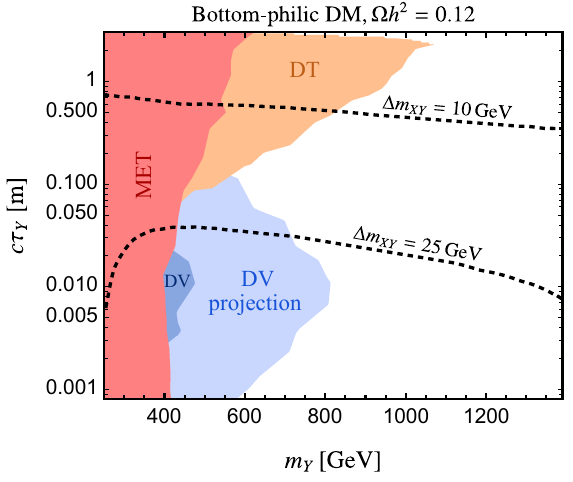} \hspace{2ex} \includegraphics[width=0.46\textwidth, trim= {0.0cm 0.0cm 0.0cm 0.0cm}, clip]{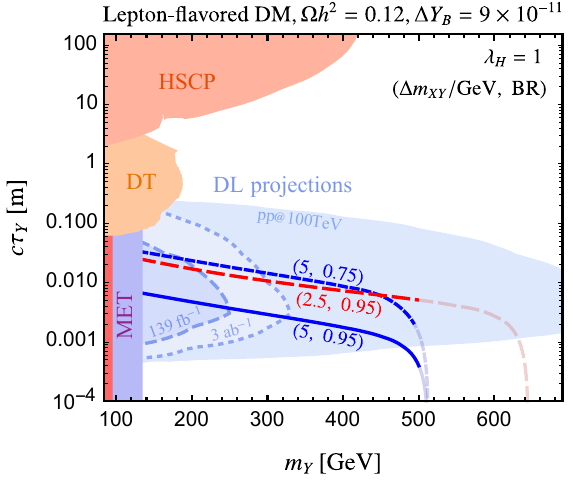} \vspace{-1.5ex} 
\caption{
Collider constraints on the cosmologically viable parameter space in conversion-driven freeze-out scenarios. Left panel: Results for a bottom-philic minimal  $t$-channel mediator model from~\cite{Heisig:2024xbh}. Right panel: Results for a lepton-flavored DM model from~\cite{Heisig:2024mwr}, which simultaneously explains the baryon asymmetry. The shaded areas labeled `MET,' `DT,' `DV,' and `HSCP' correspond to the 95\% CL exclusions from current LHC searches for missing energy, disappearing tracks, displaced vertices, and heavy stable charged particles, respectively. The light blue shaded regions labeled `DV projection' and `DL projections' indicate the potential reach of proposed searches for displaced vertices and leptons, respectively. See text and Refs.~\cite{Heisig:2024xbh,Heisig:2024mwr} for details.
}
\label{fig:paramsp}
\end{figure}

The left panel of Fig.~\ref{fig:paramsp} shows the interplay of LHC constraints on the bottom-philic minimal model from the latter three searches, specifically, the  CMS disappearing track (DT', orange shading) searches~\cite{CMS:2018rea,CMS:2020atg}, the ATLAS displaced vertex search~\cite{ATLAS:2017tny} (DV', dark blue shading), and the CMS missing energy search\cite{CMS-EXO-20-004} (MET', red shading). Remarkably, the reach of the displaced vertex search barely exceeds that of the missing energy search, despite the significantly lower background in long-lived particle searches. This highlights a gap in the current search strategies, which generally lack sensitivity to soft displaced objects. Existing searches are typically optimized for hard displaced objects, as found in scenarios like split supersymmetry. However, in conversion-driven freeze-out, the visible objects from mediator decays are typically soft -- of the order of the mass splitting, up to tens of GeV. 

In the case of the ATLAS displaced vertex search~\cite{ATLAS:2017tny}, the signal is largely suppressed due to a cut on the invariant mass of the displaced jet. Notably, a small reduction of this cut could significantly enhance the search reach, as demonstrated in Ref.~\cite{Heisig:2024xbh} and shown by the light blue shaded region (`DV projection') in the left panel of Fig.~\ref{fig:paramsp}. This is especially relevant for the upcoming LHC run, as the sensitivity of long-lived particle searches improves more substantially with increasing statistics than the one of missing energy searches, that are limited by systematics already~\cite{Heisig:2024xbh}.

The right panel of Fig.~\ref{fig:paramsp} presents the collider constraints on the lepton-flavored DM model. The curves correspond to three slices of parameter space with fixed mass splitting $\Delta m_{XY}$ and mediator branching ratio (BR) into $X_1$, each simultaneously explaining the DM relic density and baryon asymmetry. The viable parameter space lies in the millimeter to centimeter range for $c\tau$. Notably, missing energy searches~\cite{ATLAS:2019lng,CMS:2024gyw} impose the strongest constraints in this regime, however, their reach does not extend beyond mediator masses of 140\,GeV\@.

As in the quark-philic case, existing searches for displaced leptons -- potentially sensitive to these lifetimes -- lack coverage due to their focus on harder leptons. However, a dedicated search optimized for softer leptons (while additionally leveraging the missing energy from DM) is expected to significantly improve sensitivity~\cite{Heisig:2024mwr}. The corresponding light blue shaded regions indicate the potential reach of such a search at the LHC (for both $139\,\mathrm{fb}^{-1}$ and $3\,\mathrm{ab}^{-1}$) and the FCC-hh.

\section{Conclusions}\label{sec:concl}

Conversion-driven freeze-out has emerged as an appealing mechanism to explain the measured relic density with very small couplings while still providing thermalization of DM in the early Universe. The mechanism can be realized within a large class of models involving a $t$-channel mediator particle relatively close in mass to the DM particle.

Intriguingly, due to the gradual chemical decoupling of DM, it departs significantly from thermal equilibrium well before its freeze-out, while the DM abundance is still close to the relativistic one. This early departure -- satisfying Sakharov's out-of-equilibrium condition -- allows for the generation of sizeable matter-antimatter asymmetries in SM particle abundances through $CP$-violating conversion processes. Since SM asymmetries are affected by sphaleron processes while those of the new physics particles are not, a net asymmetry is frozen in without requiring any source of lepton- or baryon-number violation in the new physics sector. For a lepto-philic model, the viable parameter space -- explaining both DM and the baryon asymmetry -- extends up to DM masses of around 500\,GeV, which can be probed at (future) colliders. 

Interestingly, the required coupling for conversion-driven freeze-out coincides with a mediator decay length in the range of millimeters to meters, offering prominent signatures of long-lived particles at colliders. However, current searches at the LHC lack sensitivity toward the lower end of this range. Since the visible decay products of the mediator are typically soft, probing this scenario requires dedicated searches for soft displaced objects. This represents a gap in current LHC search strategies, making the scenario a prime target for upcoming and future improvements. Notably, the viable parameter space for conversion-driven freeze-out is bounded at large DM masses, implying full coverage at envisaged future colliders.

\providecommand{\href}[2]{#2}\begingroup\raggedright\endgroup

\end{document}